# Single-shot readout of a driven hybrid qubit in a GaAs double quantum dot

Wonjin Jang[1][†], Min-Kyun Cho[1][†], Hyeongyu Jang[1], Jehyun Kim[1], Jaemin Park[1], Gyeonghun Kim[1], Byoungwoo Kang[1], Hwanchul Jung[2], Vladimir Umansky[3], and Dohun Kim[1]*

*[1]Department of Physics and Astronomy, and Institute of Applied Physics, Seoul National University, Seoul 08826, Korea*

*[2] Department of Physics, Pusan National University, Busan 46241, Korea*

*[3]Braun Center for Submicron Research, Department of Condensed Matter Physics, Weizmann Institute of Science, Rehovot 76100, Israel*

*[†]These authors contributed equally to this work.*

*Corresponding author:* dohunkim@snu.ac.kr

**Abstract**

We report a single-shot-based projective readout of a semiconductor hybrid qubit formed by three electrons in a GaAs double quantum dot. Voltage-controlled adiabatic transitions between the qubit operations and readout conditions allow high-fidelity mapping of quantum states. We show that a large ratio both in relaxation time vs. tunneling time ($\approx 50$) and singlet-triplet splitting vs. thermal energy ($\approx 20$) allow energy-selective tunneling-based spin-to-charge conversion with readout visibility $\approx 92.6\ \%$. Combined with ac driving, we demonstrate high visibility coherent Rabi and Ramsey oscillations of a hybrid qubit in GaAs. Further, we discuss the generality of the method for use in other materials, including silicon.

Performing high-fidelity projective readout of qubit states is an important requirement in many steps of quantum information processing protocols[1–8]. In the semiconductor quantum dot (QD) qubit platform, state detection mainly uses sensors proximal to qubits, where the sensor is either sensitive to the number[9–14] or the susceptibility of the charges inside a QD to external perturbation[15–19]. Along with the progress in developing wide-bandwidth charge sensors[10,20], single-shot state detection methods have been employed in various spin qubits in semiconductors, including single-spin[21,22], singlet-triplet[23–29], and exchange-only qubits[30]. The rapidly repeated single-shot readout performed in these systems is also used for nuclear feedback control[31,32] and quantum non-demolition measurement[33,34].

The QD hybrid qubit (HQ)[35–38] compromises the desirable features of charge (fast manipulation) and spin (long coherence time) qubits. Formed by a decoherence-free subspace of three-electron spin states in a double QD, recent experiments on both Si/SiGe[36–38] and GaAs[39] have demonstrated fast electrical control of HQ with a favorable ratio between the manipulation time and coherence time $T_2^*$. Moreover, the naturally formed extended charge noise-insensitive sweet spot is tunable, and $T_2^*$ exceeding 100 ns has been demonstrated in Si/SiGe[38]. However, so far these experiments have been performed by time-averaged measurements, while more advanced protocols [1,2,4–8] using HQ require high-fidelity single-shot readout. Time-averaged measurements are also often susceptible to errors in relaxation time $T_1$ compensated probability normalization.

In this work, we demonstrate high-fidelity single-shot measurements of a three-electron HQ in GaAs. The logical states $|0\rangle$ and $|1\rangle$ are mapped to spin states that are energetically separated by more than 20 times the thermal energy, and the energy-selective tunneling (EST) events between one of the QDs and the reservoir is measured by a radio-frequency single-electron transistor (rf-set). Similar to Ref.[37], we use resonant driving to coherently control the HQ states and demonstrate high-visibility, normalization-free two-axis control on the HQ Bloch sphere. Achieving measurement fidelity $\approx 96.4$ %, readout visibility $\approx 92.6\%$, and quantum oscillation visibility $\approx 75\%$, the result facilitates efficient HQ state detection with fidelity in line with the state-of-the-art EST single-shot detections achieved in various semiconductor gate-defined QD qubits[22,25].

Fig. 1a shows a scanning electron microscope image of a QD device similar to the one we measured. The device is designed to form up to four QDs used for qubits, but we focus on the right double QD by grounding the irrelevant gate electrodes. Au/Ti gate electrodes are deposited on top of a GaAs/AlGaAs heterostructure, where a 2D electron gas (2DEG) is formed 70 nm below the surface. The device was operated in a dilution refrigerator with base temperature $\approx 20$ mK and at zero external magnetic fields. The electron temperature is $\approx 234$ mK (see SI Section S3).

A previous study of HQ in GaAs double QD showed that operating the HQ near the (2,3)-(1,4) charge occupation provides energy tunability stemming from asymmetric and anharmonic potentials[39]. Instead, we operate our HQ with the same total number of electrons as proposed originally near the (2,1) – (1,2) charge transition. We define the qubit states at the readout window as $|0\rangle = |\downarrow\rangle|S\rangle$ and $|1\rangle = \sqrt{1/3}|\downarrow\rangle|T_0\rangle - \sqrt{2/3}|\uparrow\rangle|T_-\rangle$ where $|\downarrow\rangle$ and $|\uparrow\rangle$ represent the spin configuration of the single electron in the left QD and, $|S\rangle, |T_0\rangle$, and $|T_-\rangle$ represent the singlet (*S*) and triplet ($T_0, T_-$) spin configurations of the two electrons in the right QD as in the original HQ design[35,36]. Note that the spin states comprise $|S_{tot} = 1/2, S_z = -1/2\rangle$ subspace. We describe the detailed energy levels and toy-model Hamiltonian in supporting information.

We perform ac-driven spectroscopy of the qubit frequency. As shown in the right panel of Fig. 1b, we start with an initial qubit state at the (1,2) ground level at the initialization and measurement point I/M. After adiabatically ramping the detuning $\varepsilon$ to the operation point O, the resonant ac-modulation in the detuning induces the probability to be in the excited state, $P_1$, which is adiabatically mapped back to the point I/M. The point I/M is chosen so that the Fermi level of the right reservoir resides between the energies of $|0\rangle$ and $|1\rangle$ to enable EST. The same technique was used for HQ in Si/SiGe in time-averaged probability measurement[37]. Here, we monitor the charge difference using fast rf-reflectometry recording tunneling events at MHz bandwidth, which enables single-shot projective readout. As we show below in detail, the double QD used in this work exhibits a highly asymmetric singlet-triplet splitting between the dots, where the splitting in the left (right) dot, $\delta L$ ($\delta R$) is ~ 3 (96) $\text{h} \cdot \text{GHz}$. The exceptionally small $\delta L$ may be a possible evidence for the non-negligible electron-electron interaction which is known to cause quenching of excited orbital energy spectrum[40,41] (see SI Section S2 for preliminary theoretical calculation). From the magnetic field susceptibility measurement (see SI section S2) we show that the (2,1) qubit states split by $\delta L$ have the same $S_{tot}$, and $S_z$ where the spin-conserving tunnel-coupling ensures the (2,1) ground states with $\left| S_{tot} = 1/2, S_z = -1/2 \right\rangle$ can be prepared via the adiabatic passage discussed above. Thus we interpret the (2,1) qubit states observed in this work as the HQ states, and use the toy-model Hamiltonian identical to the original HQ proposal[35,42] to simulate the energy dispersion. The calculation agrees well with the measured energy dispersion (black dashed curve, Fig. 1b). While further studies including the exact diagonalization calculation[41,43] are required to reveal the actual origin of the asymmetry, we focus on the single-shot readout of the HQ in this work and leave the detailed investigation of the energy levels for the future works.

Fig. 1c shows a double-dot charge stability diagram where the scanning gate voltage is superimposed with a voltage pulse with a rise time of $100 \text{ ps}$ and width of $10 \text{ ns}$ (schematic in Fig. 1c, see SI for zoom-out version of the diagram, and the high-frequency setup), which induces a non-adiabatic coherent Landau-Zener tunneling. The range of the gate voltage $V_2$ where these oscillations appear can be used for estimating singlet-triplet splitting $\approx 0.39 \text{ meV}$ using the measured lever arm 0.028. This is $\approx 20$ times larger than the thermal energy $\approx 20 \text{ } \mu\text{eV}$. As shown in Fig. 1d, the real-time traces of the rf-set signal at I/M show a clear distinction between $\left| 0 \right\rangle$ and $\left| 1 \right\rangle$. An electron occupying an excited orbital state of $\left| 1 \right\rangle$ tunnels to the reservoir to form the (1,1) charge state, leading to an abrupt change in the sensor signal, and initializes back to the energetically favorable $\left| 0 \right\rangle$. In contrast, no tunneling occurs for the state $\left| 0 \right\rangle$.

We analyze the performance of the single-shot readout by optimizing various tunneling rates and signal integration times. Fig. 2a depicts the time-resolved tunnel-out events, which predominantly involve triplet states, triggered at the end of the pulse sequence. We measure the tunneling-out time $\tau_{out} \approx 2 \, \mu s$ extracted from the exponential fitting. Similar measurements for tunneling in events, which occurs mostly by singlet states, result in a tunneling-in time of $\tau_{in} \approx 32 \, \mu s$ (Fig. 2a, inset). Highly asymmetric tunneling times stem from different spatial distributions of the orbital wave functions of the singlet and triplet states that lead to different dot-to-reservoir coupling[23,25]. We note that this large difference in state-dependent tunneling rates can, in principle, be used for tunneling rate-based single-shot measurement, which can be useful for reducing measurement times[23], but here we focus on the Elzerman-type readout[21] and set the measurement window to $140 \, \mu s$ longer than $\tau_{in}$. Compared with these time scales, $T_1$ at the point I/M shown in Fig. 2b, which is obtained by measuring the decay of the oscillation visibility as a function of the waiting time at point W (see inset to Fig. 2b), is longer than $100 \, \mu s$ leading to $T_1 / \tau_{out}$ about 50. Fig. 2c, which depicts the signal histogram with $1 \, \mu s$ integration time, shows a separation of the mean value of the $|0\rangle$ and $|1\rangle$ signal levels by more than 5 times the standard deviation. Using these parameters, we estimate the measurement fidelities for $|0\rangle$, $|1\rangle$, and readout visibility that accounts for measurement errors owing to relaxation and thermal tunneling events[22,25] (see SI section S4). As shown in Fig. 2d, the measurement fidelity (visibility) reaches 96.4% (92.6%) at the optimum threshold, confirming high-fidelity single-shot readout of the HQ states (see SI section S5). Moreover, using master equation simulations and additional $T_1$ measurements, we estimate that the readout error due to leakage and state relaxation during the adiabatic ramp pulse is less than 2 % (see SI section S4).

We now discuss the application of the single-shot readout method to ac-driven coherent operations of the HQ. Applying bursts of ac detuning modulation at the point O yields Rabi oscillations corresponding to $x$-axis rotations on the Bloch sphere, as shown in Figs. 3a-b. The typical Rabi frequency, which is of the order of $100 \text{MHz}$, increases linearly as a function of the microwave amplitude $A_{\text{mw}}$ at the output port of the waveform generator. Although the readout visibility with perfect gate control can be as high as $92.6\%$, the limited Rabi decay time due to decoherence and finite pulse length (see SI Section S4) leads to maximum oscillation visibility of approximately $75\%$ in this experiment.

Moreover, $T_2^*$ is characterized by performing a Ramsey experiment (Fig. 3c), which demonstrates $z$-axis rotations on the qubit Bloch sphere. Between the first and second rotation pulses $X_{\pi/2}$, which initialize the superposition state and set the measurement axis, respectively, we apply a ramp-evolution pulse with detuning amplitude $\varepsilon_{\text{P}}$. Z-axis rotation during the evolution time $t_e$ results from the development of a relative phase between $|0\rangle$ and $|1\rangle$, given by, $\varphi = -t_{\text{e}}(2\pi f_{\text{Qubit}})$ where $f_{\text{Qubit}}$ is the qubit frequency. Typically, $T_2^*$ is of the order of $7 \text{ ns}$, which is similar to earlier results (Fig. 3c, inset)[39]. While a recent theory provides coherence analysis of HQs in both GaAs and Si [44], more work is necessary for systematically identifying the dominant sources of noise in this system.

Furthermore, Fig. 3d demonstrates two-axis controllability on the *x-y* plane of the Bloch sphere. The $P_1$ oscillations of the states initially prepared along and opposite to the *y*-axis ($P(|Y\rangle)$ and $P(|-Y\rangle)$) are out-of-phase as a function of the phase $\phi$ of the measurement pulse $\Omega(\phi)_{\pi/2}$ that determines the angle between the *x*-axis and the measurement axis. Together with the Rabi (*x*-axis control) and Ramsey (*z*-axis control) oscillations, the result demonstrates the full control of the GaAs HQ with single-shot readout capability.

In this experiment, the highly asymmetric singlet–triplet energy splitting, possibly originating from the electron-electron interaction[40,41], was exploited, which provided the $f_{\text{Qubit}} \approx 1.4$ GHz regime during operation in the (2,1) configuration that facilitates electronic ac-control. It also provided the $f_{\text{Qubit}} \approx 95.8$ GHz regime in the measurement configuration (1,2), which is useful for high-fidelity EST. While the technique is general and can be used for GaAs HQ in other electron occupancies as well as silicon-based HQ[37,45], further investigations are required for determining a convenient regime for both ac-control and high-fidelity measurement. In Si/SiGe, $T_1$ at the I/M point is shown to exceed 100 ms[35] which facilitates high-fidelity single-shot readout even with a room-temperature trans-impedance amplifier. Moreover, the current quantum oscillation visibility is limited by $T_2^*$ for the given tuning. While the splitting in the energy level in the operation configuration is expected to be tunable, one cannot rule out that $T_2^*$ of the HQ in GaAs in this tuning is limited by nuclear fluctuations that mix different logical states. In such a situation, reducing $df_{\text{Qubit}}/d\varepsilon$ and hence, reducing the susceptibility to charge noise by further tuning may not necessarily increase $T_2^*$. We plan to investigate the dominant source of noise by systematic tuning as well as the HQ regime for the left double QD (Fig. 1a) in the same device.

In conclusion, we have demonstrated the high-fidelity EST-single-shot readout of a driven HQ in GaAs. Achieving a measurement fidelity $\approx 96.4\%$ and readout visibility $\approx 92.6\%$, which are comparable with state-of-the-art EST single-shot detections for other types of QD qubits[22,25,46], the results set the benchmark for HQ readout performance and provide a useful demonstration that can be adopted for HQ in a more general setting. With single-shot readout on a $\mu s$ time scale, experiments involving fast Hamiltonian learning [32,47] or detecting wide-band noise spectra [47–50] using HQ, whose demonstration has so far been limited only to single-spin and singlet-triplet qubits, are also conceivable.

**Acknowledgments**

The authors thank Mark A. Eriksson, Susan Coppersmith, Mark Friesen, and Ekmel Ercan for useful discussions. This work was supported by the Samsung Science and Technology Foundation under Project Number SSTF-BA1502-03. The cryogenic measurement used equipment supported by the National Research Foundation of Korea (NRF) Grant funded by the Korean Government (MSIT) (No.2019M3E4A1080144, No.2019M3E4A1080145, and measurement at $\varepsilon$ identical to point I/M. By observing the amplitude decay of the Larmor oscillation as a function of the waiting time at the point W indicated in the inset, $T_1 = 102 \pm 6\ \mu s$ is obtained. (**c**) Histogram of the detector signal with an integration time of $1\ \mu s$. The solid curves are the histograms for the states $|0\rangle$ and $|1\rangle$ simulated using the experimentally obtained $\tau_{\text{out}}$, $\tau_{\text{in}}$, $T_1$, and the thermal tunneling probability. (**d**) Calculated fidelity and visibility as a function of the threshold level $V_{\text{threshold}}$ showing the readout fidelity for the state $|0\rangle$ ($|1\rangle$) of 95.4 % (97.3 %). The readout visibility is 92.6 % at the optimal threshold $V_{\text{opt}}$.

**Figure captions**

**Figure 1. (a)** Scanning electron microscope image of the hybrid qubit (HQ) device similar to the one used in the experiment. Green (Yellow) circles: Double (Single) quantum dot used to form an HQ (charge sensor). **(b)** Probability of the state $|1\rangle$, $P_1$, as a function of the ramp amplitude $V_{ramp}$ and applied microwave frequency $f_{MW}$, illustrating the energy dispersion of the HQ. The black dashed line shows the calculated dispersion using the Hamiltonian described in supporting information. Inset: Schematic energy levels of the HQ as a function of the energy detuning $\varepsilon$ and pulse sequence used for the spectroscopy. **(c)** Double dot charge stability diagram near the (2,1) – (1,2) charge transition spanned by $V_1$ and $V_2$. The superimposed non-adiabatic step-pulse (blue pulse diagram) results in an oscillatory detector signal near the point I/M. **(d)** Single-shot traces of the HQ. The energy-selective tunneling (EST) readout of the HQ is enabled by putting the reservoir level between the qubit splitting. EST of $|1\rangle$ results in the step-pulse signal whereas no peak occurs for $|0\rangle$.

**Figure 2. (a)** Histogram of the tunnel-out time $\tau_{out}$. Inset: Histogram of the tunnel-in time $\tau_{in}$. Exponential fits yield, $\tau_{out} = 2.04 \pm 0.03 \ \mu s$, and $\tau_{in} = 32 \pm 3 \ \mu s$. **(b)** Relaxation time $T_1$

**Figure 3.** (**a**) P$_1$ as a function of the microwave burst time $\tau_{mw}$ and amplitude $A_{\text{mw}}$ at instrument output when the resonant driving frequency $f_{\text{mw}} \approx 1.4\text{GHz}$. The bottom panel shows that the Rabi frequency $f_{\text{Rabi}}$ increases linearly with $A_{\text{mw}}$. (**b**) Rabi oscillation of $P_1$ as a function of $f_{\text{mw}}$ and $\tau_{\text{mw}}$. Inset: Line-cut at $f_{\text{mw}} \approx 1.4\text{GHz}$. (**c**) ac-Ramsey oscillation as a function of the detuning amplitude $\varepsilon_{\text{P}}$ and free evolution time $t_{\text{e}}$. Inset: Line-cut showing $T_2^* \approx 7\text{ns}$ at the $\varepsilon_{\text{P}}$ indicated by an arrow. The bottom panel shows the fast Fourier transform of the time-domain signal indicating that the spectrum is consistent with Fig. 1(b). (**d**) Projection of the initial state along (opposite to) the $y$-axis of the Bloch sphere ($P(|Y\rangle)$ blue and $P(|-Y\rangle)$ orange) to the measurement axis controlled by the phase $\phi$ of the second rotation pulse demonstrating two-axis control of the HQ qubit on the $x$-$y$ plane of the Bloch sphere. The solid lines are fittings to the sinusoidal function. The upper insets in (**a**)- (**d**) show the schematic pulse sequences used for the corresponding measurements, and the microwave bursts with the Gaussian rising / falling edge with ~ 1 ns rise/fall time are utilized. The inclusion of the Gaussian envelop leads to negligible $P_1$ for $\tau_{mw} < 2$ ns.

**Supporting Information**

The Supporting Information includes the zoom-out version of the stability diagram, preliminary full configuration interaction calculation, magnetic field susceptibility measurement, numerical model for the hybrid qubit, experimental methods, and the details of the readout fidelity analysis.

**Supporting Information**

**S1. Stability diagram measurement down to zero-electron regime**

Here we show the stability diagram in wider region to show the three-electron occupancy at the hybrid qubit operation / readout regime in this work (Fig. S1). Also, the Pauli spin-blockade (PSB) measurements at (2,0), and (0,2) charge configuration are demonstrated respectively to show the asymmetry in the singlet-triplet splitting of the left and the right dot (Fig. S2).

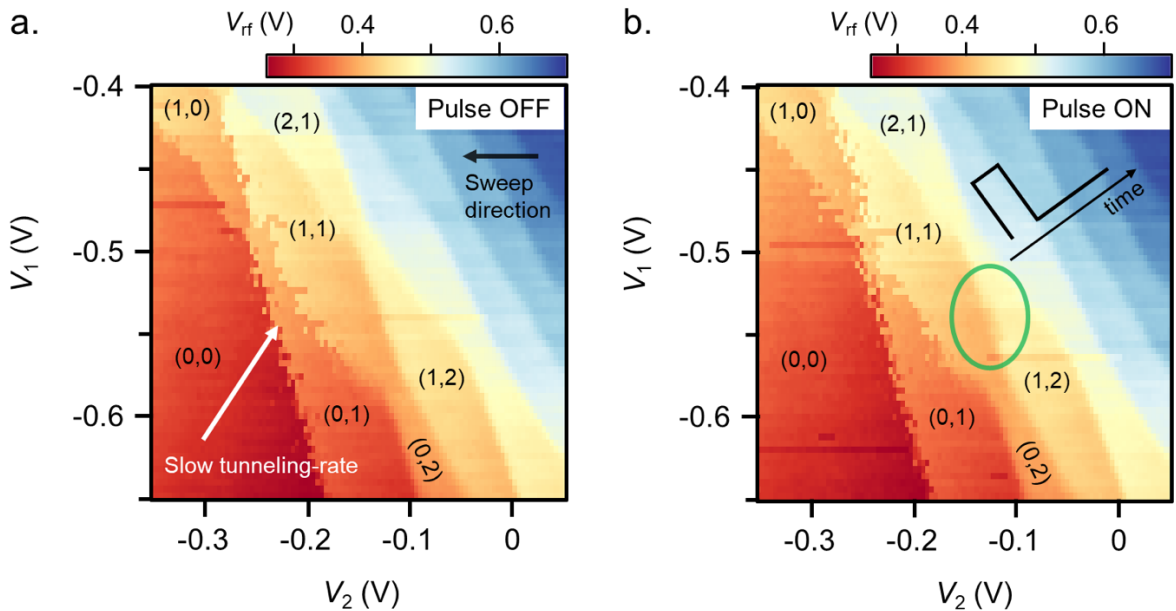

**Fig. S1. Stability diagram measurement down to zero electron regime without (a.) and with (b.) the diabatic pulse superposed. a.** Stability diagram measurement spanned by $V_1$ and $V_2$ gate voltages down to the zero electron regime confirms the three-electron occupation in the hybrid qubit readout and operation point within a double quantum dot. Due to the slow tunnel-rates in the single-electron regime charge latching behavior [1] is visible (white-arrow). **b.**

Diabatic rectangular pulse of ~ 2 ns width with ~ 33 kHz repetition rate is superposed to the dc gate voltages when measuring the same stability diagram as in a. The diabatic excitation-induced readout window is visible in the (1,2) charge regime (green circle). The sensor gate bias is compensated depending on $V_1$, and $V_2$. The $V_2$ gate voltages are swept along the direction denoted in the black arrow in a.

Fig. S2 shows the stability diagrams spanned by $V_1$, and $V_2$ near the (2,0)-(1,1) (red box), and (1,1)-(0,2) (yellow box) charge configuration. Left inset to the red {yellow} box depicts the diagram recorded by the rf-charge sensor signal where the triangular pulses with the rise-in (-out) time of ~ 20 ns (~ 300ps) toward the (1,1) charge region are superposed to the dc-voltages at ~ 20 kHz repetition rate (yellow line schematic inside the left inset). The pulse adiabatically brings (2,0)S {(0,2)S} state across the S-T+ anticrossing to (1,1)T+ state, and non-adiabatically takes the (1,1)T+ back to the (2,0) {(0,2)} region which results significant triplet population hence the PSB. The right inset to the red {yellow} box is the pulse-synced boxcar integrated signal concurrently obtained with the left inset. The boxcar integrator effectively samples and averages ~ 1 $\mu s$ signal window after the pulse non-adiabatically returns to the measurement point and reveals the PSB region as shown in the right inset to the yellow box. Note that the pulse also reveals the energy selective tunneling (EST) readout points of the $ST_0$ qubit and the hybrid qubit. In contrast, the boxcar integrated diagram near (2,0)-(1,1) charge transition does not exhibit the PSB which demonstrates the $ST_0$ splitting of the left dot is too small and PSB is lifted.

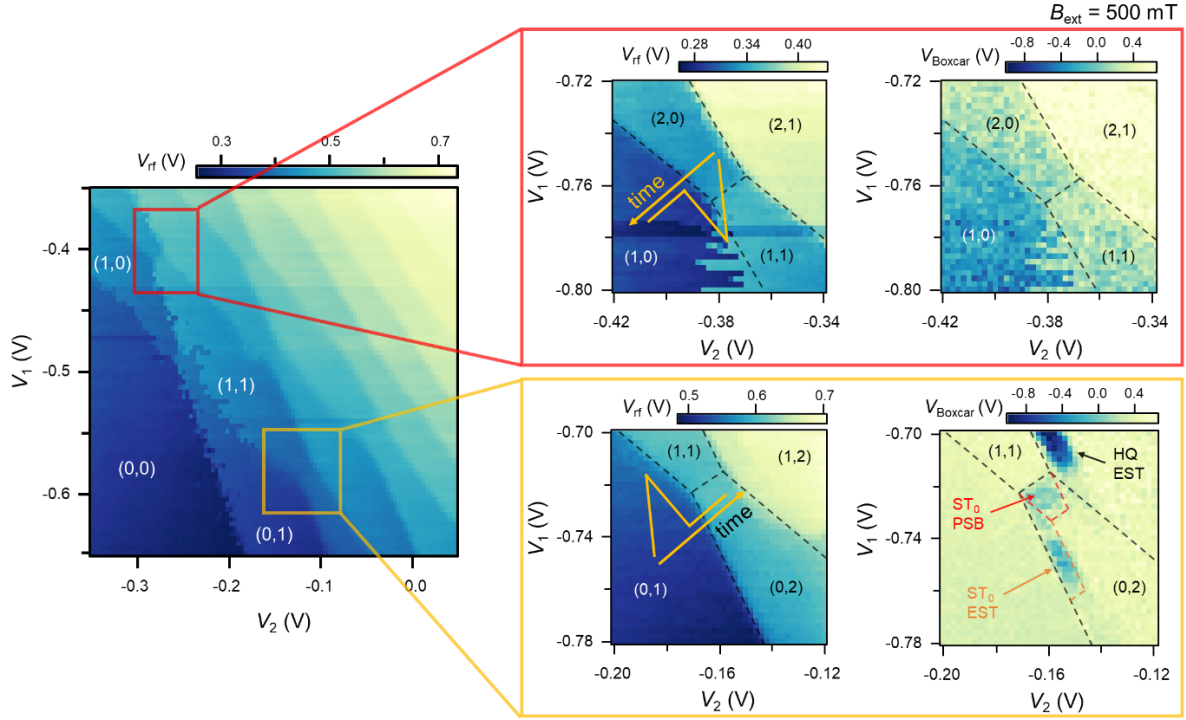

**Fig. S2. Pauli spin-blockade (PSB) measurement of the singlet-triplet (ST₀) qubit at the (2,0), and (0,2) charge configuration.** Red {yellow} box shows the stability diagrams near the (2,0)-(1,1) {(1,1)-(0,2)} charge configuration spanned by $V_1$, and $V_2$. Triangular pulse with the rise-in (-out) time of ~ 20 ns (~ 300 ps) is superposed to the dc-voltages at ~ 20 kHz repetition rate to yield the spin-blockaded (1,1) states within the (2,0) {(0,2)} charge region. Left (right) inset to both red and yellow boxes is the bare rf-charge sensor (boxcar integrated) signal. The boxcar integrator is synced to the pulse repetition rate and effectively samples ~ 1 $\mu$s window after the pulse returns to the measurement point to capture the short-lived excited state signal. The boxcar integrated signal reveals the typical PSB within the (0,2) region as well as the energy-selective tunneling readout points of the ST₀, and the hybrid qubits. In contrast, PSB is not visible in the (2,0) charge region. In-plane magnetic field, $B_{\text{ext}}$ = 500 mT is applied.

## S2. Exact-diagonalization, magneto spectroscopy, and the toy Hamiltonian

Exceptionally small singlet-triplet splitting in the left dot is unusual given that the typical size of the orbital splitting in the GaAs quantum dot (QD) is expected to be on the order of $10^1 \sim 10^2$ h·GHz [2–4]. The charge stability diagram shown in Fig. S1 confirms the double quantum dot structure, excluding the possibility of the energy modulation by the electrons from another QD. We conjecture that the electron-electron interaction which is usually not considered in the hybrid qubit (HQ) systems, is a possible reason for the extraordinary small splitting similar to recently reported works [5,6]. Likewise, here we show the preliminary calculation results derived by the full configuration interaction (FCI) method along with the electrostatic simulation and numerically demonstrate the energy splitting quenches from $\sim 10^1$ h·GHz to $10^0$ h·GHz in the left dot.

The Hamiltonian of two electrons in a quantum dot can be written as,

$$H = -\frac{\hbar}{2m^*}(\nabla_1^2 + \nabla_2^2) - eV(\mathbf{r_1}) - eV(\mathbf{r_2}) + \frac{e^2}{4\pi\varepsilon|\mathbf{r_1} - \mathbf{r_2}|} \tag{1}$$

where $m^*$ is the effective mass of the electron in GaAs, $\mathbf{r_1}$ and $\mathbf{r_2}$ are position operators, $\nabla_1^2$ and $\nabla_2^2$ are the Laplacian operators, and the confinement potential V. The spatial potential distribution V near the double-QD site is obtained self-consistently (Fig. S3a-S3c), by iteratively solving the Poisson equation within the Thomas-Fermi approximation. The distribution is calculated with the finite element method using the COMSOL Multiphysics software using the real device geometry and parameters.

As the direct diagonalization of the given two-electron Hamiltonian calls for solving ~ $10^{10}$ x $10^{10}$ sized dense matrix, we here harness the FCI method to approximate the system [5,7]. In the FCI calculation, linear combinations of all possible Slater determinants, configuration state function (CSF), were used to estimate the energy eigenstates of the multi-electron system. The CSF derived from the combination of the single-electron eigenstates are classified by the number of excited electrons. The two-electron Hamiltonian is diagonalized in the basis constructed with the ground state CSF, singly-excited CSFs, and doubly-excited CSFs.

The single-electron Hamiltonian,

$$H_1 = -\frac{\hbar}{2m^*}\nabla^2 - eV(\mathbf{r}) \tag{2}$$

can be discretized as a sparse matrix with five diagonals by adopting the five-point stencil method with the Dirichlet boundary condition. As in the harmonic potential case, we assume zero-valued wavefunctions at the boundary and calculate 100 eigenstates and eigenenergies from the discretized Hamiltonian with the Python package SciPy's '*eigsh*' method [8]. Because of the spin-degeneracy, 100 spatial eigenstates correspond to 200 spin eigenstates.

By combining the 200 spin eigenstates, one ground CSF, 396 singly excited CSFs, and 19503 doubly excited CSFs are derived. With these CSFs, the matrix elements of the Hamiltonian can be calculated with the Slater-Condon rule which connects FCI Hamiltonian matrix element with one-electron integrals and two-electron integrals for each CSF. The one-electron integral is $N^2$ sized matrix with the elements

$$\langle i|h|j\rangle = \delta_{s_i s_j}\int \psi_i^*(\mathbf{r})H_1\psi_j(\mathbf{r})\mathrm{d}\mathbf{r} \tag{3}$$

where N is the number of spin basis, $\psi_i(\mathbf{r})$ and $s_i$ are the spatial wavefunction of the $i^{th}$ single electron eigenstate and spin quantum number respectively. Also, the two-electron integral, which represents the electron-electron interaction is an $N^4$ sized four-dimensional tensor whose elements are

$$\langle ij|V_{ee}|kl\rangle = \delta_{s_i s_k}\delta_{s_j s_l}\int \psi_i^*(\mathbf{r_1})\psi_j^*(\mathbf{r_2})\frac{e^2}{4\pi\varepsilon|\mathbf{r_1}-\mathbf{r_2}|}\psi_k(\mathbf{r_1})\psi_l(\mathbf{r_2})d\mathbf{r_1}d\mathbf{r_2} \qquad (4)$$

Computation of both one and two-electron integrals are accelerated with the graphics processing units (GPUs, NVIDIA RTX 3090) using the Python library CuPy [9].

Figure S3d shows the result of the numerical calculation. Energy levels without electron-electron interaction are obtained by assigning 0 to every two-electron integral term. Since calculated coefficients of the ground state and near-ground excited states are concentrated on CSF of small excited single wavefunctions, the calculation with the 100 spatial basis sets is reasonable. According to the calculation, when the electron-electron interaction is neglected the energy splitting between the ground and the first excited state is ~ 50 h·GHz (Fig. 2d, black lines) which well matches with the typical $ST_0$ splitting reported in the previous reports [2–4]. Also, ~ 50 h·GHz splitting is smaller than the $ST_0$ splitting of the right QD as can be expected from the physical size of the QD (Fig. S3b), also showing the validity of the calculation. When the electron-electron interaction is taken into account, the energy splitting quenches down to ~ 1.3 h·GHz (Fig. S3d, blue lines) which agrees with the experimental observation. Note that this calculation is rather preliminary as the current calculation considers only the single-dot energy levels (left dot) and does not yet consider the finite nuclear field effects nor the detuning dependence of the energy levels. We do not mean to precisely fit the exact value of experimentally observed singlet-triplet splitting in the left QD. Nevertheless, the initial numerical calculation shown here successfully explains the bulk part of the physics.

At this early stage, we believe significant ellipticity of the confinement potential found in the left dot (Fig. S3b) compared to the right dot as well as a rather shallow potential depth are critical to have the dominant effect of electron-electron interaction. Roughly the shallow potential leads to the Wigner parameter $R_W = (e^2/\kappa l_0)/\hbar\omega_0$ about 5 for the left dot showing that the left dot is in the strongly correlated regime, where $\kappa$ is the dielectric constant of the GaAs, $l_0$ is the spatial extent of the 1s orbital, and $\omega_0$ is the characteristic energy splitting when the confinement potential is approximated to be parabolic. In comparison, the more strongly confined right dot (see Fig. S3b) does not experimentally show significant excited-level quenching, and part of the reason is the sufficiently negative voltage applied to the rightmost gate to tune the right dot to reservoir tunnel rate slow enough to enable single-shot readout, which leads to a more circular and symmetric dot.

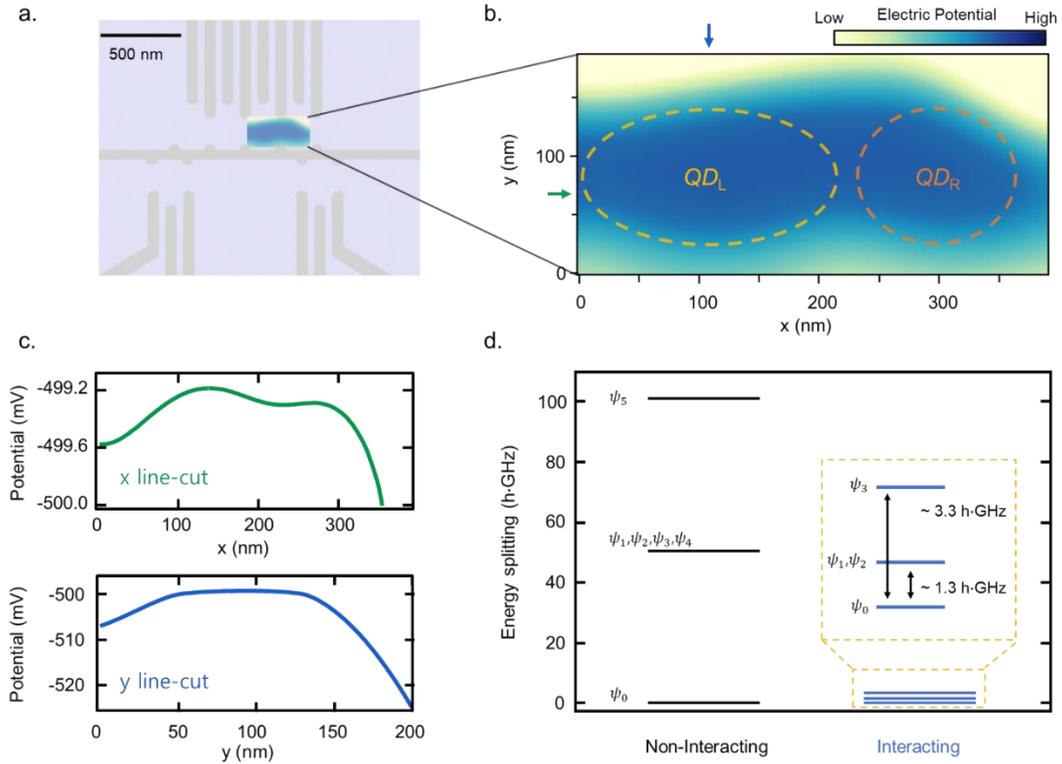

**Fig. S3. Energy splitting calculation based on Full Configuration Interaction (FCI) method. a.** Scaled gate geometry of the QD device used in this work for electrostatic simulation. Electric potential near the double-QD site is simulated by the COMSOL Multiphysics software with the dc-voltages used in the experiment. Semi-classical electron number inside the double quantum dot (integral of the Thomas-Fermi electron density over the quantum dot area) resultant from this gate voltage set was confirmed to be (2,1) **b.** Spatial distribution of the confinement potential near the QD sites. Dashed circles denote the expected position of the QDs where the left QD is expected to be distributed over an oval-shaped area. **c.** Line-cut of the potential along the x (y) direction along the green (blue) arrow in b. **d.** Diagram of the energy splitting with (blue lines) and without (black lines) the electron-electron interaction by the FCI calculation. When the electron-electron interaction is not considered, the energy difference between the ground and the first excited state is ~ 50 h·GHz. The electron-electron interaction quenches the spectrum, resulting in ~ 1.3 h·GHz difference between the ground and the first excited state.

To confirm that it is nevertheless reasonable to assume that the qubit levels behave as a QD hybrid qubit, we measured the magnetic field dependence of the energy splitting. By performing the ac-driven energy spectroscopy at a fixed detuning, we show that the energy splitting has no significant dependence on the magnetic field (Fig. S4). This implies the qubit states at the (2,1) configuration have the same $S_z$ and $S_{tot}$. Based on the observation along with the adiabatic initialization process described in the main text, we interpret the qubit states at the operation regime as the HQ states with $\left| S_{tot} = 1/2, S_z = -1/2 \right\rangle$.

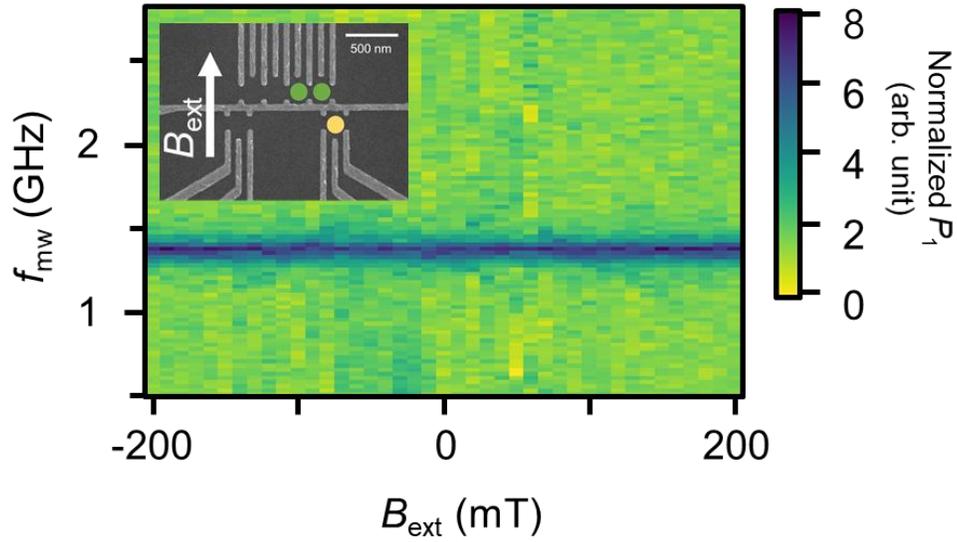

**Fig. S4 Magnetic field dependence of the energy splitting.** Microwave spectroscopy at the qubit frequency ~ 1.4 GHz is performed as a function of the external magnetic field strength, $B_{ext}$. The magnetic field is applied along the direction shown in the white arrow in the inset.

Following the HQ level spectroscopy demonstrated in Si/SiGe [10], we write the toy model Hamiltonian as,

$$H = \begin{pmatrix} \varepsilon/2 & 0 & t_1 & -t_2 \\ 0 & \varepsilon/2 + \delta L & -t_3 & t_4 \\ t_1 & -t_3 & -\varepsilon/2 & 0 \\ -t_2 & t_4 & 0 & -\varepsilon/2 + \delta R \end{pmatrix} \qquad (5)$$

The basis set for the Hamiltonian is $\{(2,1)g, (2,1)e, (1,2)g, (1,2)e\}$ where the $g$ and $e$ denote the ground and the excited state respectively at each charge configuration. The n (m) in the (n, m) notation indicates the number of electrons in the left (right) quantum dot and $\varepsilon$ is the energy detuning between the dots. $\delta L$ ($\delta R$) is the singlet-triplet energy splitting in the left (right) dot, and $t_i$ (i = 1, 2, 3, 4) denotes the tunnel coupling between the different charge states. Figure. S5a shows the eigen-energy diagram as a function of $\varepsilon$ calculated with the parameter values of $\delta L/h = 3$ GHz , $\delta R/h = 95.8$ GHz , $t_1/h = 1.8$ GHz , $t_2/h = 7.1$ GHz , $t_3/h = 11.5$ GHz , $t_4/h = 6.3$ GHz . These parameters are obtained by empirically fitting the theoretical spectrum to the experimentally observed energy dispersion (Fig. 1b) except $\delta R/h = 95.8$ GHz , fixed by the measured value described in the main text.

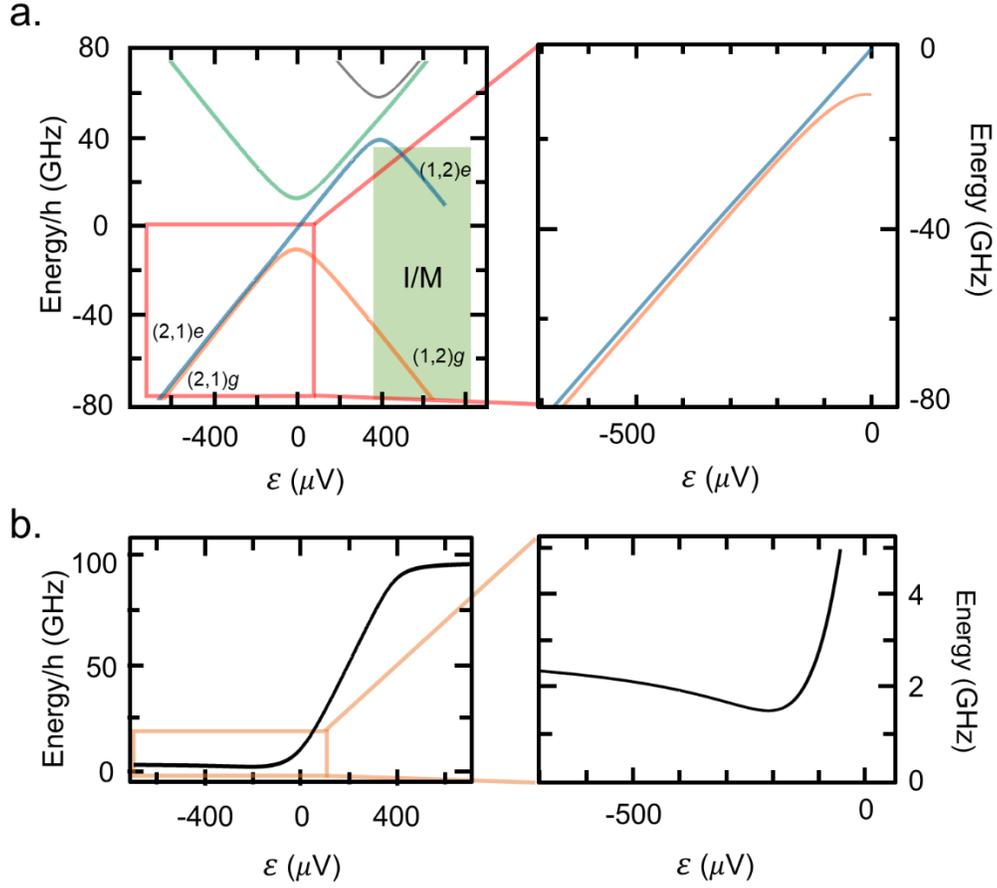

**Fig. S5. Energy level simulation a.** Eigen-energy levels of the hybrid qubit simulated with the Hamiltonian (5). The Green boxed region is the energy-selective tunneling position for qubit readout and initialization. **b.** Energy splitting between the lowest two energy levels as a function of $\varepsilon$. The right panel of **a.** (**b.**) shows the energy levels (splitting) near the operation point O of the hybrid qubit investigated in the main text.

## S3. Experimental method

The bulk of the experimental setup utilized in this work is described in Ref. [3]. The lever-arm of both $V_1$, and $V_2$ gates in Fig. 1a is 0.028 which is determined from Coulomb diamond measurements. The electron temperature $T_e \approx 234$ mK is estimated by fitting the Fermi-Dirac distribution curve to the $V_2$ electron transition line in the single electron regime (Fig. S6a).

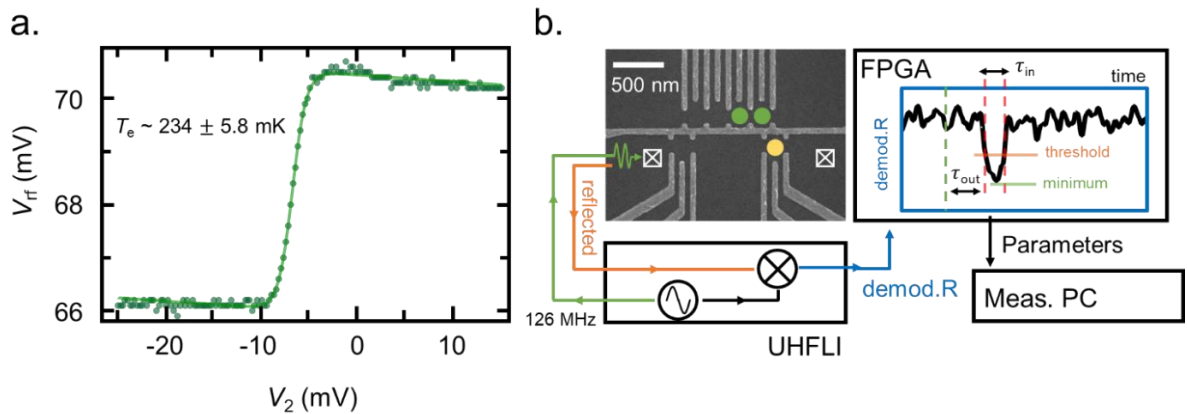

**Fig. S6. Experimental methods a.** Electron temperature measurement. By fitting the Fermi-Dirac distribution to the electron transition line, electron temperature $T_e \approx 234$ mK is extracted. **b.** Experimental setup utilized for single-shot readout of the hybrid qubit. The magnitude of the reflected rf-signal is demodulated in a high-frequency lock-in amplifier and is directly put to the field-programmable gate array (FPGA) where the signal is thresholded to discriminate the different qubit states. The FPGA also performs time-tagging of the tunneling events enabling time-resolved tunneling time measurements.

For single-shot detection, as shown in Fig. S6b, the transient tunneling events for qubit state $|1\rangle$ are thresholded in real-time using a field-programmable gate array (FPGA, Digilent Zedboard). The FPGA samples input data with the sampling rate of 1 MSa/s, compares the preset threshold with the data in parallel, and records the bit 1 immediately when data below the threshold value is detected. The bit 0 is recorded when such events did not happen throughout the preset measurement period. The sequence is repeated 5000 times to estimate the probability of the state $|1\rangle$, $P_1$. The FPGA also tags time after the trigger for each detected tunneling event, and the statistics of the tunneling-out and -in times are gathered to build the histograms shown in Fig. 2 of the main text.

For generating the high-frequency signals at the room-temperature, a high-speed arbitrary waveform generator (AWG, Keysight technologies, M8195A) which supports up to 65 GSa/s sampling rate is utilized. To combine the high-frequency and dc-signals for the gate electrodes, commercial off-board bias-tees (Tektronix, PSPL 5546) which supports > 10 GHz ac-signals are used.

With the high-frequency setup shown above, we demonstrate the coherent charge qubit Larmor oscillation [11] to evaluate the actual signal delay at the gate electrodes. Fig. S7a depicts a scanning electron microscope image of a different GaAs quantum dot device used for the charge qubit measurement, which is placed in the same cryostat with the same high-frequency wiring and circuit board as used in this work. As shown in Fig. S7a, pulses with the opposite polarity, and with the fixed width of 60 ps are generated by two different channel outputs of the AWG to be applied to $VP_1$ and $VP_2$ gate electrodes respectively. Sweeping the pulse amplitudes at the AWG output as a function of the relative delay between the two outputs of the AWG reveals a V-shaped coherent oscillation pattern (Fig. S7b). This is because the maximum detuning modulation condition (Fig. S7b, dashed line) is sensitive to the actual delay at the gate electrodes on the order of ~ 20 ps, directly implying the rise-times as short as ~ 20 ps can be transferred to the gate electrodes without further distortion.

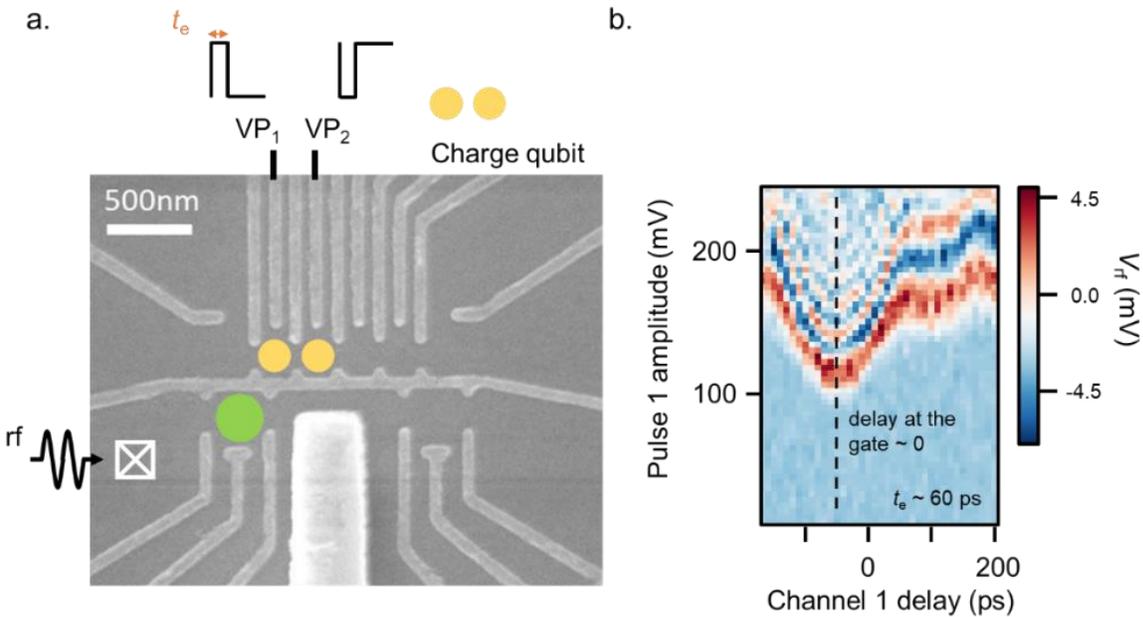

**Fig. S7. High-frequency transmission line calibration. a.** Scanning electron microscope image of the device used for the charge qubit measurement and cryostat transmission line calibration. Two square pulses of the width, $t_e$ ~ 60 ps with the opposite polarity are applied to $VP_1$ and $VP_2$ gate electrodes respectively. The pulses are generated by two different channels in an arbitrary waveform generator (AWG). rf single-electron transistor (rf-set, green dot) detects the charge states of the double quantum dot (yellow dots). **b.** Sweeping the pulse amplitude versus the relative channel delay reveals a V-shaped oscillation pattern. At ~ -50 ps channel delay (dashed line) the delay at the gate electrodes vanishes, ensuring the full 60 ps evolution at the target detuning set by the pulses.

**S4. Effects of the ramp-in and ramp-out pulse on the measurement fidelity**

We discuss the $T_1$ relaxation time at the operation regime in (2,1) charge configuration, and the effect of the ramp pulse on the read-out visibility. Fig. S8a shows detuning dependent mapping of $T_1$ times measured with pulse sequence depicted in the inset to Fig. S8a. The minimum $T_1$ time of 20 ns occurs near the detuning amplitude $\varepsilon_P \approx -170$ mV in the charge qubit regime, and the increasing $T_1$ time with respect to charge qubit energy splitting ($\varepsilon_P > -170$ mV) is consistent with the typical trend observed both in GaAs [2] and Si charge qubits [12] dominated by charge noise-induced relaxation. In the HQ regime ($\varepsilon_P < -170$ mV), $T_1$ more rapidly increase away from the anti-crossing showing reduced susceptibility to charge noise.

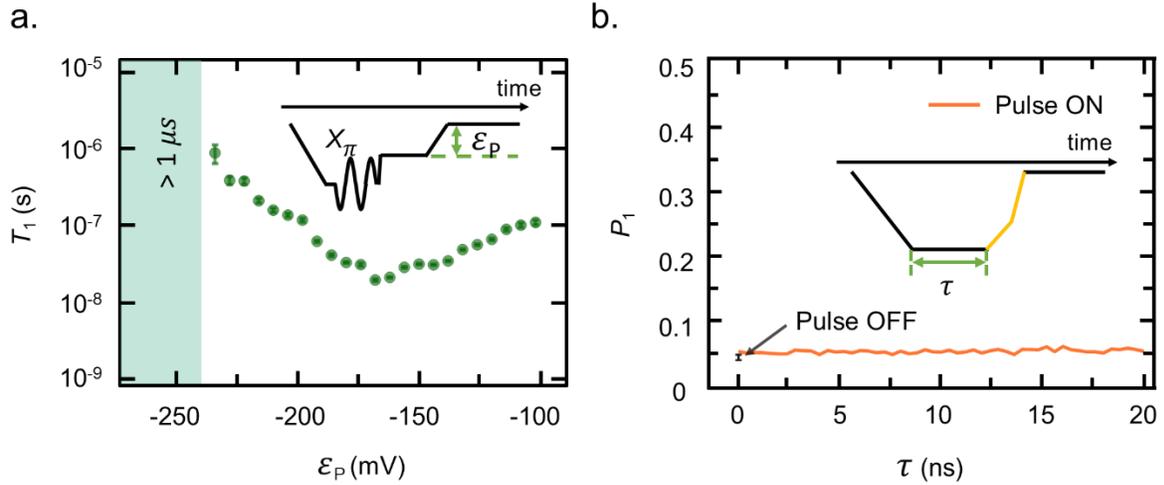

**Fig. S8. Detuning dependent $T_1$ times and effects of ramp pulses. a.** $T_1$ times measured in the qubit operation regime as a function of the detuning amplitude $\varepsilon_P$. Inset: Schematic pulse

sequence used for the $T_1$ measurement. **b.** $P_1$ as a function of dwell time $\tau$ of the adiabatic ramp-only pulse depicted in the inset showing negligible change compared with the pulse turned-off. $P_1 \sim 0.04$ is due to thermal tunneling events by the state $|0\rangle$ within the $140\mu s$ long measurement window.

The measured $T_1(\varepsilon)$ is used for investigating the effect of adiabatic ramps, during which probability leakage or energy relaxation can in principle lead to visibility loss. As shown in the inset to Fig. S8b, the two-stage ramp-out sequence is utilized to avoid the relaxation hot-spot in charge qubit regime but maintain the adiabaticity. By solving the master equation with the toy-model Hamiltonian (5) given in S2, we confirm that the leakage probability during the ramp-in stage is kept below 0.1 %. Accumulated state relaxation probability during the ramp-out near the relaxation hot-spot is $\approx 1\%$. Moreover, a relatively fast second ramp-out stage (rise time 2 ns) results in an unintentional Landau-Zener transition probability of less than $\approx 1$ %.

Fig. S8b shows the experimentally observed $P_1$ as a function of dwell time $\tau$ of the adiabatic ramp-only pulse. Although independent measurement of the state leakage or non-adiabaticity probability is challenging, the result shows that readout errors due to unintended non-adiabatic state transition during ramp or state leakage out of computational qubit states are less than 2 % consistent with the calculation. Due to the thermal tunneling probability of the state $|0\rangle$ for the given measurement duration of $140 \, \mu s$, about 4% offset is measured even when the pulse is turned off, and this effect is included in the measurement fidelity analysis in the main text and in S5. We expect that there is room for improvement to reduce the unwanted state $|0\rangle$ tunneling by applying adaptive adjustment of readout duration by further FPGA implementation and including fast initialization pulse sequence using, for example, the observed relaxation hot-spot in the charge qubit regime (Fig. S8a).

## S5. Readout fidelity analysis

Following the Ref. [13] the single-shot traces are numerically generated with the experimentally obtained parameters. By randomly assigning the qubit states to traces according to a parameter $p_1$ which corresponds to the probability for qubit $|1\rangle$ state, 8,000 single-shot traces are generated. For the case of the $|1\rangle$ state if a random variable, $p_r$, in the range [0,1] is larger than the relaxation proability calculated from the $T_1$ relaxation time and the total measrement time, a rectangular tunneling peak which follows the statistics set by the tunneling-in / -out times (Fig. 2a) is generated. In case of the $|0\rangle$ state, if a random variable, $p_t$, in the range [0,1] is smaller than the thermal tunneling probability obatined in Fig. S8b, a tunneling peak is generated. For each $|0\rangle$ and $|1\rangle$ trace, random gaussian noise is added and the numerical low pass filter similar to the experiment is applied. By sampling the minimum value from each trace, the histogram of the minimum values can be obtained to be compared with the experimentally acquired histogram. After the filtering process, $p_1$ and the gaussian noise amplitude are optimized to fit the simulated histogram to the experimental curve.

Because the information on the spin state for each simulated trace is given in priori, separate histograms corresponding to $|0\rangle$ and $|1\rangle$ states can be acquired respectively. From the separate histograms, measurement fidelity for $|0\rangle$, $F_0(V_t) = \int_{V_t}^{\infty} n_0(V) \mathrm{d}V$ and $|1\rangle$, $F_1(V_t) = \int_{-\infty}^{V_t} n_1(V) \mathrm{d}V$ are evaluated along the threshold voltage $V_t$, where the $n_0$ ($n_1$) corresponds to the normalized histogram of the $|0\rangle$ ($|1\rangle$). By choosing the optimal threshold that maximizes the visibility, $v(V_t) \equiv F_0(V_t) + F_1(V_t) - 1$ the $|0\rangle$ ($|1\rangle$) measurement fidelity 95.4 % (97.3 %), and the visibility 92.6 % are obtained.

## Supplementary references